\documentclass[conference]{IEEEtran}
\IEEEoverridecommandlockouts
\usepackage{cite}
\usepackage{amsmath,amssymb,amsfonts}
\usepackage{algorithmic}
\usepackage{graphicx}
\usepackage{textcomp}
\usepackage[table]{xcolor}
\def\BibTeX{{\rm B\kern-.05em{\sc i\kern-.025em b}\kern-.08em
    T\kern-.1667em\lower.7ex\hbox{E}\kern-.125emX}}

\usepackage{booktabs}
\usepackage{subcaption}
\usepackage{multirow}
\graphicspath{{./images/}}
\newcommand{\bs}{\boldsymbol}
\newcommand{\Var}{\mathrm{Var}}
\newcommand{\E}{\mathrm{E}}
\definecolor{Gray}{gray}{0.95}
\newcolumntype{g}{>{\columncolor{Gray}}c}    
\begin{document}

\title{Orthogonal variance-based feature selection for intrusion detection systems
\thanks{© 20XX IEEE.  Personal use of this material is permitted.  Permission from IEEE must be obtained for all other uses, in any current or future media, including reprinting/republishing this material for advertising or promotional purposes, creating new collective works, for resale or redistribution to servers or lists, or reuse of any copyrighted component of this work in other works.}
}

\author{\IEEEauthorblockN{Firuz Kamalov}
\IEEEauthorblockA{\textit{Department of Electrical Engineering} \\
\textit{Canadian University Dubai}\\
Dubai, UAE\\
firuz@cud.ac.ae}
\and
\IEEEauthorblockN{Sherif Moussa}
\IEEEauthorblockA{\textit{Department of Electrical Engineering} \\
\textit{Canadian University Dubai}\\
Dubai, UAE \\
smoussa@cud.ac.ae}
\and
\IEEEauthorblockN{Ziad El Khatib}
\IEEEauthorblockA{\textit{Department of Electrical Engineering} \\
\textit{Canadian University Dubai}\\
Dubai, UAE \\
ziad.elkhatib@cud.ac.ae}
\and
\IEEEauthorblockN{Adel Ben Mnaouer}
\IEEEauthorblockA{\textit{Department of Computer Engineering} \\
\textit{Canadian University Dubai}\\
Dubai, UAE \\
adel@cud.ac.ae}
}

\maketitle

\begin{abstract}
In this paper, we apply a fusion machine learning method to construct an automatic intrusion detection system. Concretely, we employ the orthogonal variance decomposition technique to identify the relevant features in network traffic data. The selected features are used to build a deep neural network for intrusion detection. The proposed algorithm achieves 100\% detection accuracy in identifying DDoS attacks. The test results indicate a great potential of the proposed method.
\end{abstract}


\section{Introduction}
The ubiquitous connectivity of the modern world has yielded tremendous gains. Network-based technologies allow to control and synchronize operations of complex systems. Today's networks permeate every facet of daily life making them vulnerable to unwanted intrusions and attacks. The stakes for network protection are as high as ever. It is estimated that malicious network attacks cause billions of dollars in annual damage. Network attacks can also cause physical damage as illustrated by a recent attempt to access a water treatment facility's network in Florida and pump dangerous amounts of sodium hydroxide in the water supply. Therefore, it is imperative to have effective safeguarding mechanisms that can protect networks.
The traditional intrusion detection systems (IDS) are manually designed and coded by domain experts. Although this approach served well in the past, the increased variety and volume of malicious attacks has made it more difficult to keep up with the pace. The modern IDS require automation and scalability. There is a need for rapid design and deployment of IDS to provide timely response to the fast evolving network attacks.
Our goal is to present a new approach for constructing IDS based on machine learning methods. The proposed method is shown to achieve a 100\% detection accuracy making it a promising tool in the fight against malicious attacks.

The recent advances in machine learning techniques and increased computing power have made it an attractive tool for constructing intelligent IDS. Machine learning and artificial intelligence (AI) are already being used in a range of applications against unwanted online intrusions. Random forests and neural networks have been used to build intelligent email filtering systems.  AI-based filtering algorithms have the ability to independently learn to distinguish between regular and junk messages. Intelligent systems are used to automatically discover and block suspicious URLs, add connection exceptions, and create new rules. Deep neural networks are also used to detect digital virus signatures.
One of the main advantages of AI-based IDS is scalability. Given the tremendous computing power that is available on cloud platforms such as Amazon Web Services, AI-based systems are able to handle any increase in network traffic.  AI algorithms do not require human input and can learn and evolve as necessary on their own accord.  In fact, AI algorithms benefit from increased traffic as it provides more data for learning. The processing power of AI-based IDS cannot be matched by human experts. Thus, AI offers an attractive avenue for developing automated IDS. It seems increasingly likely that machine learning will play a crucial role in protecting our networks from malicious attacks.

In this paper, we propose a novel machine learning algorithm for detecting malicious attacks. The proposed method is based on a feature selection technique using orthogonal variance decomposition \cite{Saltelli}. The algorithm is implemented in two stages. First, the network traffic data is analyzed using the orthogonal variance decomposition and the most relevant features are selected. The features are evaluated based on the total sensitivity index. Second, a deep neural network is constructed based on the selected features to identify malicious traffic in the data. 
The proposed approach provides a number of benefits:
\begin{itemize}
\item It identifies the key features in network traffic data that can help IT experts better protect the networks. 
\item It provides an effective AI-based intrusion detection system. 
\end{itemize}
The proposed algorithm can be set up to regularly train itself on new traffic data. As a result, it can remain continuously up-to-date. The test results show that the proposed method achieves a high detection rate.

One of the main challenges in feature selection is feature interactions. Relationships between features can affect their relevance with respect to the target variable. For instance, a feature that is important on its own may lose its relevance when combined with another feature if the two features are correlated. Conversely, a pair features that are unimportant individually can be highly potent when combined into a single subset. To account for all the feature interactions, in theory, every possible subset of the feature set must be considered. Since the total number of subsets is equal to $2^k$, where is $k$ is the number of features, it is a computationally infeasible task. Despite various attempts to address the problem of feature interactions it remains an open problem \cite{Zuech}. The orthogonal variance decomposition provides a method for including all the feature interactions under the assumption of feature independence. The variance of the target variable is decomposed according to the features. Consequently, a feature responsible for a higher proportion of the target variance is deemed more relevant.
Although feature independence is a seemingly stringent requirement, the method has been shown to perform well even when this condition is not met \cite{Kamalov3}.

Network traffic data is notoriously imbalanced. The majority network signals consists of regular traffic with only a small portion representing malicious attacks. Imbalanced data leads to bias in many machine learning classifiers \cite{Thabtah}. Since the main goal of a classifier is to maximize the overall accuracy, it focuses on the majority data at the expense of the minority data. There are exists a number of approaches to combat imbalanced data \cite{Kamalov1}. The proposed approach is capable of handling imbalanced data by selecting the most relevant features.

Our paper is structured as follows. In Section 2, we present an overview of the existing literature on the subject matter. In Section 3, we describe the details of the proposed approach for constructing AI-based IDS. Section 4 contains the results of the numerical experiment to measure the performance of the method. We conclude with brief remarks in Section 5

\section{Literature review}

The recent advances in machine learning have propelled their application in a range of domains. In particular, it has been  used successfully in IDS. A comparison of popular machine learning techniques for IDS application including SVM, random forest, and extreme learning machine (ELM)  was conducted in \cite{Ahmad}. The results of numerical experiments showed that ELM outperformed other approaches. Machine learning based IDS have yielded mixed results. 
Intrusion detection in a cloud-based environment is investigated in \cite{Salman}. The authors tackle both anomaly detection and categorization of network attacks. The proposed approach  employs a combination of linear regression and random forest algorithms. The results reveal 99\% detection and 93.6\% categorization accuracy.
The authors in \cite{Kasongo} employ feature selection that is based on a boosting algorithm to reduce the dimensionality of data space. Then a number of machine learning algorithms are applied on the reduced dataset. The results show the effectiveness of the feature selection algorithm in increasing the detection rate up to 90.85\%.
A multi-faceted approach based on optimizing the model training size was proposed in \cite{Injadat}. The authors are able to reduce the computational time while achieving 99\% detection accuracy. 
In \cite{Fang}, the authors employ a combination of Elman neural network and SVM to develop a method for  intrusion information detection. The proposed method achieves detection rate between 87.3-100\%

Feature selection is used to reduce the number of variables and obtain a simpler model. It is an important preprocessing step in a number of machine learning applications \cite{Kamalov4, Thabtah2}. In particular, feature selection has been  applied in the context of IDS. In \cite{Amiri}, the authors propose two separate forward search selection algorithms based on different criteria: linear correlation coefficient and mutual information. Alternatively, in \cite{Kamalov2}, the authors propose to merge mutual information and mutual correlation into a single metric. The resulting feature evaluation method is combined with the decision tree algorithm to predict DDoS attacks. The authors in \cite{Ambusaidi} employ mutual information-based feature selection combined with SVM to design an IDS. One of the main challenges in feature selection is feature interactions. To address this issue, a new method based on orthogonal variance decomposition was proposed in \cite{Kamalov3}. The proposed methods takes into account all the feature interactions in decomposing the variance of the target variable under the condition of pairwise feature independence.

\section{Variance decomposition and intrusion detection}
The proposed algorithm consists of two parts: feature selection and deep learning. In the first step, we apply orthogonal  variance decomposition to identify the relevant features in network traffic data. The contribution of each feature in the total  variance of the target variable can be quantified in terms of the total sensitivity index (TSI). Features are subsequently scored based on the TSI. The second step of the algorithm involves training an artificial neural network on the traffic data using the features selected in the preceding step. The neural network architecture consists of several fully connected layers and a binary output. The network hyperparameters are tuned using cross validation. 
For convenience, we refer to the proposed algorithm as the total sensitivity neural network (TSNN). 

The primary purpose of orthogonal variance decomposition is to decompose the variance of the target variable in terms of the feature variables. Furthermore, variance decomposition takes into account feature interactions. Concretely, the variance of the target variable $Y$ is decomposed as 
\begin{equation}\label{eq_c}
V(Y) = \sum_i V_i + \sum_{i, j} V_{ij} + ... + V_{12 .. k}, 
\end{equation}
where each term $V_{i_1 i_2 .. i_s}$  represents the contribution to the variance of $Y$ due to feature interactions in subset $\{X_{i_1}, X_{i_2},... X_{i_s}\}$.
To obtain the decomposition in Equation \ref{eq_c}
suppose that the target variable $Y$ is a function of a set of feature variables $Y = f(X_1, X_2, ...X_k)$.  Let the features $\{X_i\}$ be independently and uniformly distributed over the interval $[0, 1]$.
Then we obtain the following functional decomposition
\begin{equation}\label{eq_fun}
f = f_0 +\sum_i f_i + \sum_{i, j} f_{i j} + ... + f_{12 .. k},
\end{equation}
where
$f_0 = \E[Y],
f_i(x) = \E[Y|X_i=x] - f_0,
f_{i  j}(x,y) = \E[Y|X_i=x, X_j=y] - f_i(x) -f_j(y) -f_0
$
and similarly for higher orders. Note that $\E[f_{i_1 i_2 .. i_s}] = 0.$ To obtain the variance decomposition, we square and integrate the two sides of Eq. (\ref{eq_fun}) 
$$\int_{[0, 1]^k} f^2\, d\boldsymbol{X} = \int_{[0, 1]^k} \big(f_0 +\sum_i f_i + \sum_{i, j} f_{i j} + ... + f_{1 2 .. k}\big)^2  \, d\boldsymbol{X},$$
where $\boldsymbol{X}$ is the vector of $X_i$'s and $[0, 1]^k$ is the $k$-dimensional hypercube.
The decomposition in Equation \ref{eq_c} can be used to determine the total contribution of an individual feature to the target variance. This is done by calculating the Total Sensitivity Index (TSI) of a feature given by 
\begin{equation}
\label{eq_tsi}
S_{T_i} = 1 - \frac{\Var(\E[Y|\bs{X}_{\sim i}])}{V(Y)},
\end{equation}
where $\bs{X}_{\sim i}$ is the vector of all features except $X _i$. The details of Equations \ref{eq_c} and \ref{eq_tsi} can be found in \cite{Kamalov3, Saltelli, Sobol}.
 
There exists a number of estimators for $\Var(\E[Y|\bs{X}_{\sim i}])$. 
We follow the approach in \cite{Homma}.
Let $\bs{A}$ and $\bs{B}$ be a pair of independent sampling matrices. Let $j$ and $i$ denote row and column indexes respectively. Define $\bs{A_B}^{(i)}$ to be matrix $\bs{A}$, where its $i$th column replaced with the $i$th column of $\bs{B}$. Then the variance estimator is given by
\begin{equation}\label{eq_estim}
\Var(\E[Y|\bs{X}_{\sim i}]) = \frac{1}{n}\sum_{j=1}^n f(\bs{A})_j f(\bs{A_B}^{(i)})_j -f_0^2 
\end{equation}
The details of the algorithm for implementing the final TSI calculations can be found in \cite{Kamalov3}. Although the utilized feature selection algorithm is more advanced most of the existing methods its implementation and complexity are reasonable. There exist many algorithms for calculating TSI that can be used for implementation.

\begin{figure}[!htb]
\centering
\includegraphics[clip, trim=14cm 4cm 8.5cm 4cm, width=0.5\textwidth]{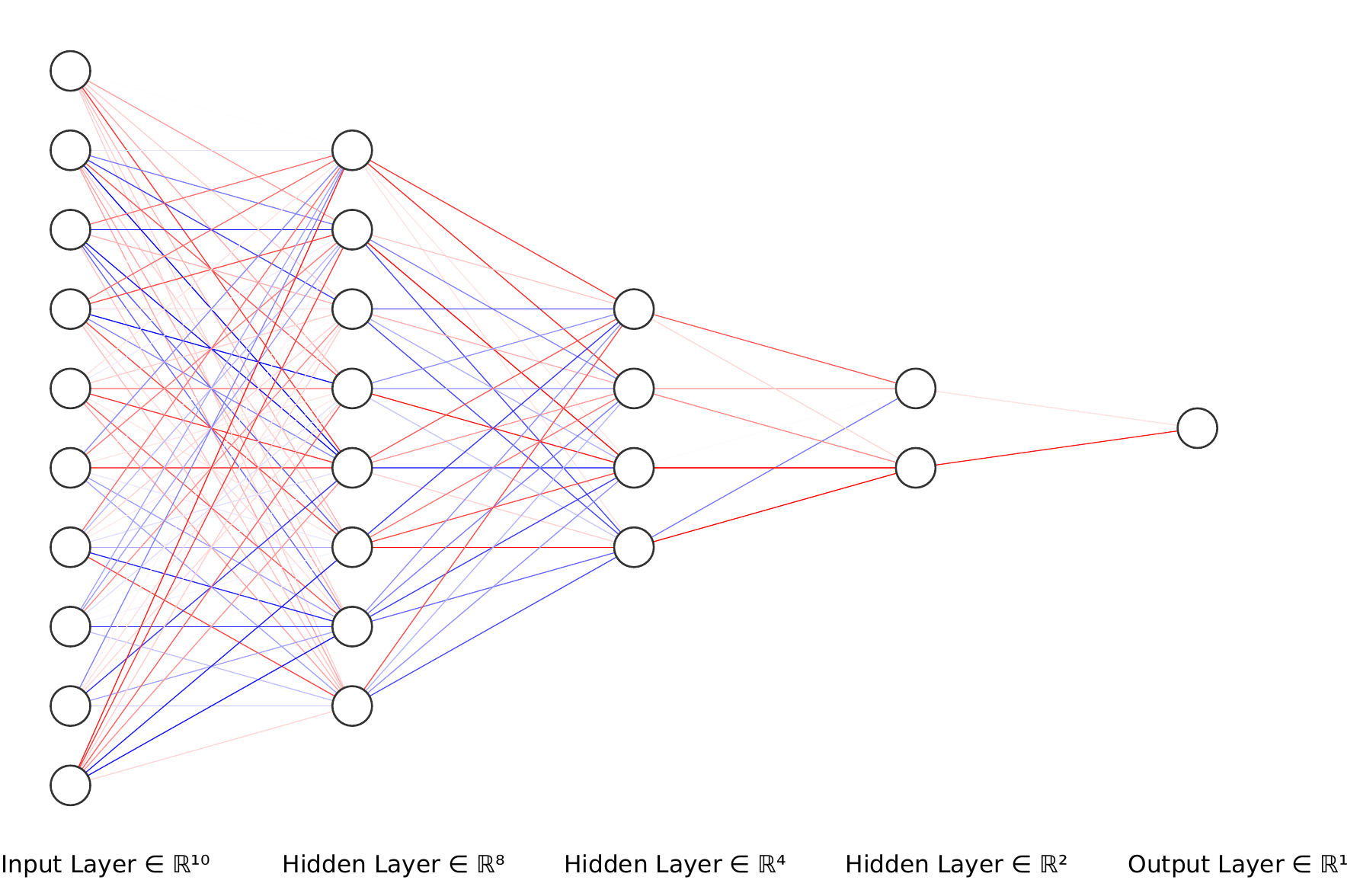}
\caption{The ANN architecture used to classify benign and malicious attacks. For illustration purposes, the edge opacity and color are proportional to edge weights. Note that in the actual model the input size is 63.}
\label{ann}
\end{figure}

After selecting the relevant features, we employ an artificial neural network (ANN) to classify the network traffic as malicious or benign. Neural networks have achieved success various domains including image and speech recognition which has prompted their application in IDS. Neural networks have the ability to learn nonlinear hidden patterns in data through several layers of abstraction.
As shown in Figure \ref{ann}, the proposed ANN is constructed with 5 fully connected layers, the size of the ANN was chosen after experimenting with different size architectures. Concretely, experiments with 3 and 4-layer architectures yield slightly lower accuracy than the 5-layer ANN.
We use $L_2$-regularization with $\lambda=0.00001$ to prevent overfitting. The ReLU activation function is used to introduce nonlinearity in the ANN model. The hyperparameters are tuned using cross validation. Binary crossentropy was used as the loss function to train the ANN.

The TSNN model can be updated on a regular basis to keep up with new threats. We propose a weekly retraining of the model based on newly available data about DDoS attacks. The decomposition step in the TSNN can be performed by an expert or automatically based on a set of predetermined criteria. To deal with different types of attacks the proposed method can be trained on a larger dataset that includes samples various DDoS instances. Alternatively, multiple TSNN models can be trained for different types of DDoS attacks.
\section{Methodology and results}
In this section, we describe the numerical experiments that were conducted to test the performance of the proposed intrusion detection algorithm TSNN. All the computations including the implementation of the feature selection method and the neural network were done in Python using Keras \cite{Chollet} and scikit-learn \cite{Pedregosa} libraries. 
\subsection{Data}
\label{data}
The data in the experiments was obtained from the collaborative project between the Communications Security Establishment and the Canadian Institute for Cybersecurity \cite{Sharafaldin}. It is based on the creation of user profiles  containing abstract representations of events and behaviors seen on the network. The dataset is generated by simulating LAN network topology frequently found on the AWS computing platform. For our purposes, we extracted a random sample of size 6,000 from the original dataset. The data is distributed according to 5/1 ratio between benign and DDoS instances. The class distribution represents a realistic scenario where the majority of signals consist of regular network traffic.
Each instance of the dataset consists of 63 continuous features and a label. The features consists of various packet and IAT characteristics. The dataset is split into training and test subsets according to 80/20 ratio. The training subset is used for feature selection, ANN training and validation while the test set is used to measure the unbiased performance of the algorithm.

\subsection{Benchmarks}
We benchmark the performance of the proposed TSNN algorithm against two commonly used machine learning algorithms: support vector machines (SVM) and logistic regression (LR). The SVM classifier is a maximum margin classifier. It is designed to find the separating hyperplane with the largest margin between the two target classes. Its main advantage is the ability to handle nonlinear tasks by mapping the data into a higher dimensional representation space using the kernel trick. Logistic regression is a simple linear classifier that maximizes the log-likelihood probability of the sample data. Its main advantage is the low variance which results in less overfitting. The two benchmark classifiers are trained on the full feature set while the TSNN is trained only on a subset of features. Despite the greater amount of information available for training the benchmark algorithms, the TSNN method yields better results.
\subsection{Results}
We test the efficacy of the proposed TSNN algorithm on the dataset described in Section \ref{data}.
First, we apply the algorithm to select the top 10 most relevant features. The selected features are presented in Table \ref{tab:top}. Among the selected features \textit{ACK Flag Coun, Init\_Win\_bytes\_forward}, and \textit{PSH Flag Count} are the most important. 
The ACK Flag Count parameter belongs to the flag group of attributes.  It represents the number of times the ACK flag bit is set to 1 for a given flow of packets sent in forward and backward directions. The flag group of attributes can be used to extract backscattered packets that result from a spoofed denial-of-service attack where the victim responds to a spoofed IP address of another victim used in the DDOS attack instance. The common headers of TCP packets used in flag attributes are SYN+ACK, RST, RST+ACK, and ACK \cite{Fachkha}.
The min\_seg\_size\_forward and Init\_Win\_bytes\_forward features belong to the flow descriptors group. They are useful for volumetric-flow monitoring and show the quantity of packets in a given direction, minimum, maximum, the mean and standard deviation of the packets as well as other descriptive statistics. These features are used to monitor and detect volume-based DDoS attacks that use a large amount of malicious traffic to bring down a resource. 

\begin{table}[h!]
\centering
\caption{The top 10 features selected by the features selection method.}
\label{tab:top}
\begin{tabular}{lrl}
\toprule
{} &     TSI &                  feature \\
\midrule
\rowcolor{Gray}
1  &  0.2361 &           ACK Flag Count \\
2  &  0.0847 &   Init\_Win\_bytes\_forward \\
\rowcolor{Gray}
3  &  0.0847 &           PSH Flag Count \\
4  &  0.0713 &         act\_data\_pkt\_fwd \\
\rowcolor{Gray}
5  &  0.0653 &                 Idle Std \\
6  &  0.0535 &     Avg Bwd Segment Size \\
\rowcolor{Gray}
7  &  0.0490 &        Max Packet Length \\
8  &  0.0446 &   Packet Length Variance \\
\rowcolor{Gray}
9  &  0.0431 &   Total Backward Packets \\
10 &  0.0416 &      Subflow Bwd Packets \\
\bottomrule
\end{tabular}
\end{table}

 
In the second step of the algorithm, we utilize a deep neural network to classify benign and malicious network signals based on the selected features in Table \ref{tab:top}. We split the data into training ans test subsets according to 80/20 ratio. The ANN is trained for 200 epochs with batch size 500 using the root mean square propagation optimizer. The summary of the results is presented in Table \ref{summary}. The test results show accuracy of 100\% in distinguishing between the benign and DDoS traffic signals. Since the model accuracy is calculated based on the test set - independent of the training set - the chances of overfitting are minimal. Our result compares favorably with similar results in the literature. In addition, we trained and tested the benchmark methods SVM and LR which produced accuracy rate of 98.41\% and 99.67\% respectively. In addition, the TSNN method achieves perfect performance in precision and recall. Note that the benchmark methods were trained on the full dataset using all 63 features. Despite utilizing fewer features, the TSNN algorithm outperformed the benchmark methods. 

\begin{table}[h!]
\centering
\caption{Summary of the experiments.}
\label{summary}
\begin{tabular}{lrrrr}
\toprule
{Algorithm} & \# features   &  Accuracy rate & Precision & Recall\\
\midrule
TSNN  &     10  & 100\%		& 100\%  		& 100\% \\
SVM &   63 &	98.41\%  &	98.86\%	&	91.05\%\\
LR  & 63		&	99.67\%	&	100\%	&	97.89\%\\
\bottomrule
\end{tabular}
\end{table}

One the main issues with the modern machine learning methods is interpretability. Algorithms that are trained on large datasets with a big number of features produce black-box models that are often impossible to humanly comprehend. Therefore, reducing the number of features required to build a model provides a valuable advantage. Models with fewer number of features such as TSNN have better interpretability.

\section{Conclusion}
The advent of ubiquitous network based technologies has increased the associated vulnerabilities. The need for effective network protection tools has never been greater. In this paper, we propose an AI-based IDS that is capable of distinguishing between regular and DDoS traffic. The proposed method fuses an advanced feature selection technique together with deep learning to produce a simple yet efficient IDS. The proposed algorithm produces 100\% accuracy on the tested dataset.

Despite the encouraging results there remains more work to be done to improve the proposed algorithm. Issues such as encrypted payload must be addressed. Other variables such as flow features and packet headers should also be considered.
Although additional numerical experiments are required to validate the proposed method, the initial results offer a promising avenue for further research.

\end{document}